\documentclass
[12pt]{article}
\usepackage{amssymb}
\usepackage{epsfig}

\catcode`\@=11
\def\numberbysection{\@addtoreset{equation}{section}
        \def\theequation{\thesection.\arabic{equation}}}

\def\beq{\begin{equation}}
\def\eeq{\end{equation}}
\numberbysection
\begin{document}
\begin{titlepage}
\begin{center}
\hfill  \\
\vskip 1.in {\Large \bf Remarks on cosmological gravitational waves} \vskip 0.5in P. Valtancoli
\\[.2in]
{\em Dipartimento di Fisica, Polo Scientifico Universit\'a di Firenze \\
and INFN, Sezione di Firenze (Italy)\\
Via G. Sansone 1, 50019 Sesto Fiorentino, Italy}
\end{center}
\vskip .5in
\begin{abstract}
We study the propagation of a gravitational wave in an AdS spacetime. We find that, in presence of the cosmological constant, the graviton mass cannot be measured with higher precision than
$\sqrt{\Lambda}$.
\end{abstract}
\medskip
\end{titlepage}
\pagenumbering{arabic}
\section{Introduction}

The discovery of gravitational waves is undoubtedly an important confirmation of general relativity \cite{1}. Because gravity describes our universe at large scales, it is possible to use this new experimental tool to test general properties of the universe. Among the various possible measures \cite{2} we mention, for example, the presence of the cosmological constant. The physical motivations for studying the cosmological gravitational waves have been widely discussed in \cite{3} and we do not repeat them here. Our purpose is to solve the Einstein equations in the Lorentz gauge in order to clarify from a theoretical point of view the interaction between the cosmological constant and the gravitational wave solution. The article is organized as follows.
First we perturbatively develop the Einstein equations with cosmological constant up to the second order, leaving undifferentiated the perturbation of the metric $ h^{(2)}_{\mu\nu} $. We show that the equation of motion for the metric $ h^{(2)}_{\mu\nu} $ satisfies the Lorentz gauge, provided that $ h^{(1)}_{\mu\nu} $ is constructed to satisfy the same gauge condition. Then we split the perturbation of the metric $ h^{(2)}_{\mu\nu} $ into three parts, $ h^{(2) \Lambda}_{\mu\nu}, h^{(2) G}_{\mu\nu}, h^{(2) \Lambda G} _{\mu\nu} $. To see the interaction between the $ AdS $ space-time \cite{4} and the gravitational wave it is necessary to study the eq. of motion for the mixed term $ h^{(2)\Lambda G}_{\mu\nu} $, which is linear in the gravitational wave solution. Then we analyze the eq. of motion for $ h^{(2)\Lambda G}_{\mu\nu} $ with the Fourier transform and we provide the simplest solution, which is however already indicative of the type of solution we can expect.

\section{Einstein equations}

The propagation of a gravitational wave in an $AdS$ space-time can be described by an Einstein manifold of the form

\beq R_{\mu\nu} \ = \ \Lambda \ g_{\mu\nu} \label{21} \eeq

which solves the Einstein's equations. Similar to the case $ \Lambda = 0 $, we can look for perturbative solutions for the metric of the type:

\beq g_{\mu\nu} \ = \ \eta_{\mu\nu} \ + \ h_{\mu\nu}^{(1)} \ + \ ..... \label{22}
\eeq

with the Lorentz gauge condition

\beq  \eta^{\mu\mu'} \partial_{\mu'} h_{\mu\nu}^{(1)} \ = \ \frac{1}{2} \partial_\nu ( \eta^{\mu\mu'} h_{\mu\mu'}^{(1)} ) \label{23}
\eeq

The perturbative development probably makes sense only in certain coordinate systems that describe the $AdS$ space but not in all \cite{3}.

At the first perturbative order the equation to be solved is of the type

\beq \eta^{\mu\mu'} \partial_{\mu} \partial_{\mu'} h_{\mu\nu}^{(1)} \ = \ - \ 2 \Lambda \ \eta_{\mu\nu} \label{24} \eeq

whose solution can be split into two distinct parts:

\beq h_{\mu\nu}^{(1)} \ = \ h_{\mu\nu}^{(1)\Lambda} \ + \ h_{\mu\nu}^{(1)G} \label{25} \eeq

where

\begin{eqnarray} h_{\mu\nu}^{(1)\Lambda} & = & - \frac{\Lambda}{9} ( \ x_\mu x_\nu + 2 \eta_{\mu\nu} x^2 \ )  \ \ \ \ \ \ \ \ \ {\rm  AdS \ solution} \nonumber \\
h_{\mu\nu}^{(1)G} & = & \ {\rm  gravitational \ wave} \label{26} \end{eqnarray}

The $ h_{\mu\nu}^{(1) G} $ field satisfies the special conditions

\begin{eqnarray}  \eta^{\mu\mu'} \partial_{\mu} \partial_{\mu'} \ h_{\mu\nu}^{(1)G} & = & 0  \nonumber \\
\eta^{\mu\mu'} \partial_{\mu'} h_{\mu\nu}^{(1)G} & = &  h^{(1)G} \ = \ 0  \label{27} \end{eqnarray}

At this order there is no interaction between AdS space-time and gravitational wave.

\section{Second order perturbative development}

To study the interaction between AdS space-time and gravitational wave we must move to the second perturbative order:

\beq g_{\mu\nu} \ = \ \eta_{\mu\nu} \ + \ h_{\mu\nu}^{(1)} \ + \ h_{\mu\nu}^{(2)} + ..... \label{31}
\eeq

The connection is more complicated now

\begin{eqnarray} \Gamma_{\alpha\beta}^{ \mu (2) } & = & \Gamma_{\alpha\beta}^{ \mu (2) \ I} \ + \ \Gamma_{\alpha\beta}^{ \mu (2) \ II} \nonumber \\
\Gamma_{\alpha\beta}^{ \mu (2) \ I} & = & \frac{1}{2} \ \eta^{\mu\mu'} \ ( \ \partial_\alpha h_{\mu'\beta}^{(2) } \ + \ \partial_\beta h_{\mu'\alpha}^{(2) } \ - \
\partial_{\mu'} h_{\alpha\beta}^{(2) } \ ) \nonumber \\
\Gamma_{\alpha\beta}^{ \mu (2) \ II} & = & - \frac{1}{2} \ h^{\mu \mu'}_{(1) } \ ( \ \partial_\alpha h_{\mu'\beta}^{(1) } \ + \ \partial_\beta h_{\mu'\alpha}^{(1) } \ - \
\partial_{\mu'} h_{\alpha\beta}^{(1) } \ ) \label{32} \end{eqnarray}

The contribution of the quadratic term in $ h^{(1)} $ i.e. $ \Gamma_{\alpha\beta}^{\mu(2) \ II} $ to the curvature tensor is of the type:

\begin{eqnarray}
R_{\beta\mu\nu}^{(2) \alpha \ II} & = & \partial_\mu \Gamma_{\beta\nu}^{(2) \alpha \ II} \ - \ ( \mu \leftrightarrow \nu ) = \nonumber \\
& = & - \frac{1}{2} \ \partial_\mu h^{\alpha \alpha'}_{(1)} \ ( \ \partial_\beta h_{\alpha' \nu}^{(1)} \ + \ \partial_\nu h_{\alpha' \beta}^{(1)} \ - \  \partial_{\alpha'} h_{\beta\nu}^{(1)}
\ )  + \nonumber \\
& + &  \frac{1}{2} \ \partial_\nu h^{\alpha \alpha'}_{(1)} \ ( \ \partial_\beta h_{\alpha' \mu}^{(1)} \ + \ \partial_\mu h_{\alpha' \beta}^{(1)} \ - \  \partial_{\alpha'} h_{\beta\mu}^{(1)}
\ ) + \nonumber \\
& + &  \frac{1}{2} \ h^{\alpha \alpha'}_{(1)} \ ( \ \partial_\mu \partial_{\alpha'} h_{\beta \nu}^{(1)} \ + \ \partial_\nu \partial_\beta h_{\alpha' \mu}^{(1)} \
- \ \partial_{\mu} \partial_{\beta} h_{\alpha' \nu}^{(1)} \ - \ \partial_{\nu} \partial_{\alpha'} h_{\beta \mu}^{(1)} \ )
\label{33} \end{eqnarray}

and the contribution to Ricci tensor is

\begin{eqnarray}
R_{\beta\nu}^{(2) \ II} & = &  \delta^\mu_\alpha \ R_{\beta\mu\nu}^{(2) \alpha \ II} = \nonumber \\
& = & - \frac{1}{4} \ \partial^\alpha h_{(1)} \ ( \ \partial_\beta h_{\alpha \nu}^{(1)} \ + \ \partial_\nu h_{\alpha \beta}^{(1)} \ - \  \partial_{\alpha} h_{\beta\nu}^{(1)}
\ )  + \nonumber \\
& + &  \frac{1}{2} \ \partial_\nu h^{\alpha \alpha'}_{(1)} \  \partial_\beta h_{\alpha' \alpha}^{(1)} \ + \nonumber \\
& + &  \frac{1}{2} \ h^{\alpha \alpha'}_{(1)} \ ( \ \partial_\alpha \partial_{\alpha'} h_{\beta \nu}^{(1)} \ + \ \partial_\nu \partial_\beta h_{\alpha \alpha'}^{(1)} \
- \ \partial_{\alpha} \partial_{\beta} h_{\alpha' \nu}^{(1)} \ - \ \partial_{\nu} \partial_{\alpha'} h_{\beta \alpha}^{(1)} \ )
\label{34} \end{eqnarray}

The contribution to the Ricci tensor of $ \Gamma_{\beta\nu}^{(2) \alpha \ I} $ is instead:

\beq R_{\beta\nu}^{(2) \ I} \ = \ \delta^\mu_\alpha \ ( \ \partial_\mu \Gamma_{\beta\nu}^{(2) \alpha \ I} \ - \ ( \mu \leftrightarrow \nu )
 \ ) \ = \ - \frac{1}{2} \ \eta^{\mu\mu'} \partial_{\mu}\partial_{\mu'} \ h_{\beta\nu}^{(2)} \ + \  {\rm gauge \ terms}
\label{35}\eeq

Now let's move on to the non-linear terms

\begin{eqnarray}
R_{\beta\mu\nu}^{(2) \alpha \ III} & = &  \Gamma_{\mu\alpha'}^{(1) \alpha} \ \Gamma_{\beta\nu}^{(1) \alpha'} \ - \ ( \mu \leftrightarrow \nu ) \ = \
\nonumber \\
& = & + \frac{1}{4} \ ( \ \partial_\mu h_{\alpha'}^{(1) \alpha} \ + \ \partial_{\alpha'} h_{\mu}^{(1) \alpha} \ - \ \partial^\alpha h_{\alpha' \mu} \ )
( \ \partial_\beta h_\nu^{(1) \alpha'} \ + \ \partial_{\nu} h_{\beta}^{(1) \alpha'} \ - \ \partial^{\alpha'} h_{\beta \nu}^{(1)} \ ) - \nonumber \\
& - &  \frac{1}{4} \ ( \ \partial_\nu h_{\alpha'}^{(1) \alpha} \ + \ \partial_{\alpha'} h_{\nu}^{(1) \alpha} \ - \ \partial^\alpha h_{\alpha' \nu} \ )
( \ \partial_\beta h_\mu^{(1) \alpha'} \ + \ \partial_{\mu} h_{\beta}^{(1) \alpha'} \ - \ \partial^{\alpha'} h_{\beta \mu}^{(1)} \ ) \label{36} \end{eqnarray}

and its contribution to the Ricci tensor is

\begin{eqnarray}
R_{\beta\nu}^{(2) \ III} & = &  \delta^\mu_\alpha \ R_{\beta\mu\nu}^{(2) \alpha \ III} \ = \nonumber \\
& = & + \frac{1}{4} \ \partial_{\alpha'} h^{(1)} \ ( \ \partial_\beta h_\nu^{(1) \alpha'} \ + \ \partial_{\nu} h_{\beta}^{(1) \alpha'} \ - \ \partial^{\alpha'} h_{\beta \nu}^{(1)} \ ) \ - \
\nonumber \\
& - &  \frac{1}{4} \ \partial_\nu h_{\alpha'}^{(1) \alpha} \ \partial_\beta h_{\alpha}^{(1) \alpha'} \ + \
\frac{1}{2} \ \partial_\alpha h_{\beta}^{(1) \alpha'} \ \partial^\alpha h_{\alpha' \nu}^{(1)} \ - \
\frac{1}{2} \ \partial_{\alpha'} h_{\nu}^{(1) \alpha} \ \partial_\alpha h_{\beta}^{(1) \alpha'}
\label{37} \end{eqnarray}

Adding the various contributions we obtain the perturbative development of the Einstein equations at the second order:

\begin{eqnarray}
R_{\beta\nu}^{(2)} & = &  \Lambda h_{\beta \nu}^{(1)} \ = \ - \frac{1}{2} \ \eta^{\mu \mu'} \partial_\mu \partial_{\mu'}  h_{\beta \nu}^{(2)} \ + \
\frac{1}{4} \ \partial_\nu h_{(1)}^{\alpha \alpha'} \ \partial_\beta h^{(1)}_{\alpha \alpha'} \ + \nonumber \\
& + & \frac{1}{2} \ \partial_\alpha h_{\beta}^{(1) \alpha'} \ ( \ \partial^\alpha h_{\alpha'\nu}^{(1)} \ - \ \partial_{\alpha'} h_{\nu}^{(1) \alpha} \ ) \ +
\nonumber \\
& + &  \frac{1}{2} \ h_{(1)}^{\alpha \alpha'} \ ( \ \partial_\nu \partial_\beta h^{(1)}_{\alpha \alpha'} \ + \ \partial_\alpha \partial_{\alpha'} h^{(1)}_{\beta \nu}
\ - \ \partial_\nu \partial_{\alpha'} h^{(1)}_{\alpha \beta} \ - \ \partial_\alpha \partial_\beta h^{(1)}_{\alpha' \nu} \ )
\label{38} \end{eqnarray}

We then show that this equation of motion satisfies the Lorentz gauge condition. Let us apply the operator $ \partial_\beta $:

\beq \partial^\beta R_{\beta\nu}^{(2)} \ = \ \Lambda  \ \partial^\beta h_{\beta \nu}^{(1)} \label{39} \eeq

After many simplifications we get

\begin{eqnarray}
& \ & \eta^{\mu \mu'} \partial_\mu \partial_{\mu'} \ \partial^\beta  h_{\beta \nu}^{(2)} \ + \ 2 \Lambda \ \partial_\nu h^{(1)} \ = \ \nonumber \\
& = &  \partial^\beta h_{(1)}^{\alpha \alpha'} \ \left( \ \frac{3}{2} \partial_\nu \partial_\beta h^{(1)}_{\alpha \alpha'} \ - \ \partial_\nu \partial_{\alpha'} h^{(1)}_{\alpha \beta}
\ \right) \label{310} \end{eqnarray}

From the equation for $ h_{\beta\nu}^{(2)} $ we calculate the trace:

\beq \frac{1}{2} \ \eta^{\mu \mu'} \partial_\mu \partial_{\mu'} \ h^{(2)} \ + \ 2 \Lambda \ h^{(1)} \ = \ \frac{3}{4} \ \partial_\alpha h_{(1)}^{\alpha' \beta}
\ \partial^\alpha h^{(1)}_{\alpha' \beta} \ - \ \frac{1}{2} \ \partial_\alpha h_{(1)}^{\alpha' \beta} \ \partial_{\alpha'} h^{(1) \alpha}_\beta
\label{311} \eeq

Applying the operator $ \partial_\nu $ to this last formula we obtain:

\beq \frac{1}{2} \ \eta^{\mu \mu'} \partial_\mu \partial_{\mu'} \ \partial_\nu h^{(2)} \ + \ 2 \Lambda \ \partial_\nu h^{(1)} \ = \
\frac{3}{2} \ \partial_\beta h_{(1)}^{\alpha \alpha'}
\ \partial_\nu \partial^\beta h^{(1)}_{\alpha \alpha'} \ - \ \partial_{\alpha'}  h_{(1)}^{\beta \alpha} \ \partial_{\alpha} \partial_\nu  h^{(1) \alpha'}_\beta
\label{312} \eeq

By exchanging the $ (\alpha, \alpha', \beta) $ indices between them, we get a complete simplification:

\beq \eta^{\mu \mu'} \partial_\mu \partial_{\mu'} \ \left( \ \partial^\beta  h_{\beta \nu}^{(2)} \ - \ \frac{1}{2} \ \partial_\nu h^{(2)} \ \right) \ = \ 0
\label{313} \eeq

therefore the Einstein equations developed at the second perturbative order are compatible with the Lorentz gauge condition.

Similarly to the splitting of the perturbative solution at the first order (\ref{25}), we are going to split $ h_{\beta\nu}^{(2)} $ into three fields:

\beq h_{\beta \nu}^{(2)} \ = \ h_{\beta \nu}^{(2) \Lambda} \ + \ h_{\beta \nu}^{(2) \Lambda G} \ + \ h_{\beta \nu}^{(2) G}
\label{314} \eeq

The equation for $ h_{\beta\nu}^{(2) \Lambda} $ is determined only by the contribution $ h_{\beta\nu}^{(1)} \ \rightarrow h_{\beta\nu}^{(1) \Lambda} $ and we get the same equation
discussed in \cite{5}:

\beq \eta^{\mu \mu'} \partial_\mu \partial_{\mu'} h_{\beta \nu}^{(2) \Lambda} \ = \ \frac{44}{81} \ ( \ x_\beta x_\nu \ + \ 2 \eta_{\mu \nu} \ x^2 \ )
\label{315} \eeq

The equation for $ h_{\beta\nu}^{(2) G} $ is determined only by the gravitational wave portion of $ h_{\beta\nu}^{(1)} \ \rightarrow h_{\beta\nu}^{(1) G} $ and is the same as the vacuum Einstein equation with $ \Lambda = 0 $.

It remains to discuss the mixed contribution $ h_{\beta\nu}^{(2) \Lambda G} $ which is the result of the mixed terms of the type $ h_{\beta\nu}^{(1) \Lambda} \ h_{\beta\nu}^{(1) G} $. It is this contribution that is responsible for the physical effect of the cosmological constant on the propagation of a gravitational wave. Finally we get, using the special conditions (\ref{27}):

\begin{eqnarray}
& \ & \eta^{\mu \mu'} \partial_\mu \partial_{\mu'} h_{\beta \nu}^{(2) \Lambda G} \ = \  - \ \frac{4 \Lambda}{3} \ h_{\beta \nu}^{(1) G} \ + \
\frac{2 \Lambda}{9} \ x^\alpha \ ( \ \partial_\beta h_{\alpha \nu}^{(1) G} \ + \ \partial_\nu h_{\alpha \beta}^{(1) G} \ ) \ - \nonumber \\
& - &  \frac{2 \Lambda}{3} \ x^\alpha \ \partial_\alpha h_{\beta \nu}^{(1) G} \ - \ \frac{\Lambda}{9} \ x^\alpha x^{\alpha'} \ ( \ \partial_\nu \partial_\beta
h_{\alpha \alpha'}^{(1) G} \ + \ \partial_\alpha \partial_{\alpha'} h_{\beta \nu}^{(1) G} \ - \ \nonumber \\
& - & \partial_\nu \partial_{\alpha'} h_{\alpha \beta}^{(1) G} \ - \ \partial_\alpha \partial_\beta h_{\alpha' \nu}^{(1) G} \ )
\label{316} \end{eqnarray}

The solution of this equation gives the answer to the problem posed in this work.

\section{Fourier transform}

The equation (\ref{316}) can be analyzed with the Fourier transform. But first we need to make an approximation, i.e. to substitute the fields $ h_{\beta\nu}^{(2) \Lambda G} $ and
$ h_{\beta\nu}^{(1) G} $ with the same field $ h_{\beta\nu} $, in order to extract the leading effect. By doing so we neglect the sub-leading effects $ O (\Lambda^2) $.

We pose

\beq h_{\beta \nu} ( \vec{x} ) \ = \ c \ \int \ h_{\beta \nu} ( \vec{q} ) \ e^{ i \vec{q} \cdot \vec{x} } \ d^4 q
\label{41} \eeq

from which we obtain the equation

\begin{eqnarray}
& \ & - q^2 \ h_{\beta \nu} ( \vec{q} ) \ = \  - \ \frac{4 \Lambda}{3} \ h_{\beta \nu} ( \vec{q} ) \ - \
\frac{2 \Lambda}{9} \ \frac{\partial}{\partial q_\alpha} \ ( \ q_\beta h_{\alpha \nu} ( \vec{q} ) \ + \ q_\nu h_{\alpha \beta} ( \vec{q} ) \ ) \ + \nonumber \\
& + &  \frac{2 \Lambda}{3} \  \frac{\partial}{\partial q_\alpha} \ (  \ q_\alpha h_{\beta \nu}  ( \vec{q} ) \ )  \ - \ \frac{\Lambda}{9} \
\frac{\partial}{\partial q_\alpha} \frac{\partial}{\partial q_{\alpha'}} \ ( \ q_\nu q_\beta
h_{\alpha \alpha'} ( \vec{q} ) \ + \ q_\alpha q_{\alpha'} h_{\beta \nu} ( \vec{q} ) \ - \ \nonumber \\
& - & q_\nu q_{\alpha'} h_{\alpha \beta} ( \vec{q} ) \ - \ q_\alpha q_\beta h_{\alpha' \nu} ( \vec{q} ) \ )
\label{42} \end{eqnarray}

To solve eq. (\ref{42}) for $ h_{\beta\nu} (\vec {q}) $ we look for a solution like this

\beq h_{\beta \nu} ( \vec{q} ) \ = \ q_\beta q_\nu \ A( q^2 ) \ + \ \eta_{\beta \nu} \ B ( q^2 )
\label{43} \eeq

and we obtain the following system of equations

\begin{eqnarray}
& \ & - q^2  \ A ( q^2 ) \ = \ + \ \frac{4 \Lambda}{9} \ q^2 \ A' ( q^2 ) \ + \ \frac{4 \Lambda}{9} \ B' ( q^2 ) \ + \ \frac{4 \Lambda}{9} \ q^2 B'' ( q^2 ) \nonumber \\
& \ & - q^2  \ B ( q^2 ) \ = \  - \ \frac{4 \Lambda}{9} \ B ( q^2 ) \ - \ \frac{2 \Lambda}{3} \ q^2 B' ( q^2 ) \ - \ \frac{4 \Lambda}{9} \ q^4 B'' ( q^2 )
\label{44} \end{eqnarray}

This system allows to derive $ A (q^2) $ and $ B (q^2) $. The simplest solution is obtained by placing $ B (q^2) = 0 $, from which we finally obtain

\beq A ( q^2 ) \ = \ A ( \Lambda ) \ e^{ - \frac{9}{4\Lambda} q^2 } \ \ \ \ \rightarrow \ \delta ( q^2 ) \ {\rm \ for \ } \Lambda \rightarrow 0
\label{45} \eeq

 which is a representation of the Dirac delta distribution. We can of course look for more complicated solutions similar to those of the gravitational wave $ h_{\beta\nu}^{(1) G} $,
but we think that in any case the general characteristic is that of replacing the typical $ \delta (q^2) $ of gravitational waves (graviton mass = 0) to a representation of the Dirac delta function, with amplitude typically linked to the cosmological constant.

From this solution (\ref{45}) we deduce therefore that the mass of the graviton has an intrinsic uncertainty proportional to $ \sqrt{\Lambda} $, which constitutes a theoretical limit to its resolution in an $AdS$ space.

\section{Conclusion}

In this article we have studied the propagation of a gravitational wave in an AdS space. To obtain a valid prediction we have been forced to develop the Einstein equations up to the second order in the perturbation of the metric, imposing the Lorentz gauge. At the second order the metric perturbation $ h_{\beta\nu}^{(2)} $ can be split into three contributions,
$ h_{\beta\nu}^{(2) \Lambda} $ which is due to the pure space-time AdS, $ h_{\beta\nu}^{(2) G} $ which is purely of the gravitational wave type and $h_{\beta\nu}^{(2) \Lambda G} $
which is a mixed term. It is this last term that produces the correct interaction between the AdS space-time and the gravitational wave. After several calculations we have succeeded in extracting the linear differential equation in $ h_{\beta\nu}^{(2) \Lambda G} $ which can be solved by the Fourier transform. The Fourier analysis produces another differential equation, which can be solved. From the simplest solution we can already extract an important information. In practice the typical Dirac delta function $ \delta (q^2) $ that defines the gravitational wave (with graviton mass equal to zero) is replaced by a representation of it with amplitude proportional to the cosmological constant. This leads us to think that the main effect of the AdS space-time on the propagation of a gravitational wave is to generate an indetermination in the measure of the graviton mass equal to $ \sqrt {\Lambda} $.

\end{document}